\makeatletter\let\ifGm@compatii\relax\makeatother

\documentclass[aps,pre,floatfix,superscriptaddress,twocolumn,10pt]{revtex4-1}
\usepackage{ulem}
\usepackage{color}
\usepackage{graphicx}
\usepackage{amsmath, amsthm, amssymb}
\usepackage[colorlinks,citecolor=red,urlcolor=blue]{hyperref}
\newcommand{\be}{\begin{equation}}
\newcommand{\ee}{\end{equation}}
\newcommand{\bea}{\begin{eqnarray}}
\newcommand{\eea}{\end{eqnarray}}

\begin{document}
\title{On the relationship of Mathematics to the real world}

\author{Deepak Dhar} 
\email{deepakdhar1951@gmail.com}
\affiliation{Indian Institute of Science Education and Research, Dr. Homi Bhabha Road, Pashan,  Pune 411008, India}

\date{\today}

\begin{abstract}
   In this article, I discuss the relationship of mathematics to the physical world, and to other spheres of human knowledge. In particular, I argue  that Mathematics  is created by human beings, and the number $\pi$ can not be said to have   existed $100,000$ years ago, using  the conventional meaning of the word `exist'. 

\end{abstract}


\maketitle

{\it ``The book (nature) is written in mathematical language, and the symbols are triangles, circles and other geometrical figures, without whose help it is impossible to comprehend a single word of it; without which one wanders in vain through a dark labyrinth.'' }\\

~~~~~~~~~~~~~~~~~Galileo Galilei, in {\it Il Saggiatore} (1623).

The relationship between Mathematics  and Science, where the latter is here taken to be  the study of the real world,  has fascinated philosophers of science for a long time.  I read about some of these ideas when I was younger, of an impressionable age, and accepted them without much thought.  But in the last few years, on further reading, and ruminating about this topic,  I realized that what I took for granted as  obvious truth true then, I no longer believe in now.  I want to share my new-found wisdom with the readers. The ideas expressed are not original, and have been  discussed  by  many other people at  many other times. I am only reiterating them here, as they still  are  contrary  to the prevailing  conventional wisdom. 

In general, the view of practicing scientists is to stay away from philosophical discussions, and the advice given to young research scholars is to `shut up, and calculate.' In fact, philosophizing is considered bad manners, and a sign of old age, by many scientists. I  think that some of this disrepute may be blamed on  the opacity of many   philosophical discussions. `Philosophy is the misuse of a terminology, which was invented just for this purpose' \cite{dubislav}.

About `shut up and calculate': Many  people would agree that the so-called philosophical questions  are the more important ones.  This article is aimed at the younger readers. I do not think one should discourage the young people from thinking about philosophical questions,  just because the answers may not always be easy to find,  or because there are no  clear and unambiguous answers, or because this does not lead to a publication. Also, philosophical discussions need not be incomprehensible.  I will try to  keep the arguments here transparent.    A discussion of philosophical issues amongst young people stimulates ideas,  promote critical thinking, and may even clear misconceptions. This is my  hope. 

Just one more point. Often readers of philosophical arguments have some prior beliefs, and if the writer says something they already believe in, they go ``Right. Right. Right.'' If, on the other hand, it does not, they immediately dismiss the writer as wrong, {\it  without making any attempt to reexamine their own beliefs in the light of the arguments given}. I  hope that you will not do this.

\section{The everyman's view of Mathematics}

The popular sentiment about Mathematics is either of unadulterated hate, or of Awe and Supreme Reverence. The latter is captured in the idea  that God is a mathematician, (or takes orders from a mathematician).  One may find similar views expressed in  well-known sayings that describe  Mathematics as `the crest of a peacock'\cite{crest}, or the `Queen of Sciences'\cite{gauss}.  

The first quote is from the Vedangas, and this suggests that this is the inherited wisdom of our sages,  and was a generally accepted view at that time.  However, one may ask how many people in ancient or medieval India, at any given time, could be called mathematicians, in the sense that they at least knew about  Aryabhatta's or Bhaskara's  work, not just their names, and could explain it to others, even if they did not write any books on mathematics themselves? Would it be of the order of  $5$, or $50$, or $500$? Most educated estimates  about this number from experts tend to be nearer to 5, than to 50. Hence, it seems to me that, in spite of the quote, and in spite of the well-known  achievements of Indian Mathematics of very high order,   Mathematics  has not been held in  such a  high regard,  in practice, in the Indian philosophical tradition.  The second quote, from Gauss, presumably conveys truthfully what he  believed, but can not be considered  unbiased.  We would like to   take a harder, less starry-eyed  look at the relation of Mathematics to other spheres of human knowledge.

To be specific, let us start with the question, ``Did the number $\pi$ exist $100,000$ years ago?''  I suspect that a good fraction of readers is thinking ``Yes. Obviously!!!" What I will like to argue below is that the answer is not so obvious, if you think about it a bit.   
Of course, the answer  depends on what we  mean by `exist'.  Clearly, the number $\pi$ is not a physical object, like a table, or the planet Jupiter, and it does not exist in the same sense that a table exists. A material object has mass, and  occupies an identifiable region of space and time. A number like $\pi$ is a concept, and can exist only as such.

For example, one may speak of `eight-headed zebras'. Such  animals do not exist in the real world anywhere. But, just by putting these words together, I create these objects as a `mental category'.  It starts {\it existing} in the world of ideas.   One can then deduce several properties of such objects.  How many eyes does an eight-headed zebra have?  The answer is sixteen, as each head has two eyes.  

What is true for eight-headed zebras, is equally true for the perfect circles of Eulidean geometry. There are no perfect circles to be found anywhere in the real universe, but one can prove theorems about perfect circles from the definition, as is done in  high-school geometry text-books.

In 1960, Wigner wrote a very  influential article titled {\it ``The Unreasonable Effectiveness of Mathematics in Natural Sciences''}\cite{wigner}. This has led to a lot of discussion amongst philosophers of science. The points made by Wigner have been elaborated upon,  analyzed, and critically discussed  by many others. In particular, building on earlier discussion of Hamming, Derek Abbott, a professor of Electrical Engineering at the University of Adelaide, Australia wrote a counter-point titled ``The Reasonable Ineffectiveness of Mathematics'' in 2013 \cite{abbott}.  I found his arguments rather persuasive, and they  led me to change my position. By this article, I am trying to spread the good word.

\begin{figure}
\includegraphics[width=2.5cm,angle=0] {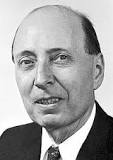} ~~~~~~~~~~~~~~~~~~~~~\includegraphics[width=2.4cm,angle=0] {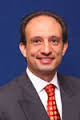}
\caption{Eugene  Wigner and Derek Abbott \cite{photos}.}
\end{figure}

\section{The Wigner Argument}
~\\
Let me start by summarizing the Wigner argument. 
Wigner starts his article with a  story about two friends, who were classmates at high school, and then meet again  after many years, and are discussing what they do. One of them has become a statistician, and shows a reprint of his latest paper on population trends.  The friend asks what is a normal distribution, and this  is  explained in terms of the  distribution of heights of men  in a city.  He asks about the symbol $\pi$ in the equation, and on being told that it is the ratio of circumference of a circle to its diameter, he is unconvinced, and thinks the friend is putting him  on, as ``surely the population has nothing to do with the circumference of the circle''. 

Wigner comments that the reaction of the classmate only betrayed common sense. He uses this example to  build his main thesis that 
``mathematical concepts turn up in entirely unexpected places. Moreover, they {\it often} permit an unexpectedly close and accurate description of the phenomena."  

He then goes on to explain the terms ``Mathematics" and  ``Physics".  It seems necessary to briefly discuss this here also, even at the risk of sounding pedantic, and  turning off some  readers.  The point is that there is no  agreement about what these words mean, and in addition, the meaning has  evolved with time. 

For example, Georg Ohm, of the Ohm's law fame,  in  his paper dealing with currents induced in wires by applying  potential differences wrote that he believed that investigations would {\it ``secure inconvertibly to Mathematics the possession of a new field of Physics, from which it had hitherto been totally excluded }.."\cite{ohm}. This  was written only about 200 years ago. But, the  sentence seems  rather strange  to a modern reader.  Ohm's law is clearly a law of Physics. Since when did it become a part of   Mathematics? Clearly, Ohm's use of the word `Mathematics' does not conform to its present meaning.  Some clarity about what we are calling `Mathematics' here is  necessary. I guess that Ohm used it in the same sense as some students do, when they say that  the equation ``$ s = 1/2 g t^2$'' is mathematics, but  the equivalent  statement  ``for falling bodies, acceleration is constant" is not. This (wrongly) identifies Mathematics with  the use of a {\it mathematical  equation} to describe the  relationship between  numerical measures of physical quantities.

Wigner's  description of Mathematics is somewhat obscure:
``... I would say that mathematics is the science of {\it skillful operations with concepts and rules,  invented just for this purpose.}" Examples of mathematical concepts he gives are complex numbers, algebras, and linear operators. He notes that  these concepts are additional ingredients to the mathematical structure.   ``A mathematician could formulate only a handful of interesting theorems without defining concepts beyond those contained in the axioms..".

~\\
About the `sciences', he notes that, in general, the world around us is unpredictable, but ``in spite of its baffling complexity, certain regularities can be observed". These are `{\it the laws of nature}', for example, the Kepler's laws of planetary motion. He says that it is a miracle, and not `natural',  that  `laws of nature' exist, and are  the same everywhere. Even more miraculous is the fact that  man is able to discover these laws. The study of these laws is what Wigner calls `Science'.

He cites the Galileo's remark given in the beginning of this article, and notes, ``A physicist uses some mathematical concepts for the formulation of laws of nature, and only a small fraction of mathematical concepts is used in Physics... Mathematical formulation of the physicist's often crude experience leads in an uncanny number of cases to an amazingly accurate description of a large class of phenomena"\\
~\\
He cites as examples the calculation of ground state energy of helium, and Lamb shift calculation of quantum electrodynamics, and concludes with `` the miracle of appropriateness of mathematics for the formulation of laws of physics is a wonderful gift that we neither understand, nor deserve. We should be grateful for it, and hope that it will remain valid in future .."

\section{The Abbott Counterpoint}
I now try to summarize the arguments of Abbott.  Abbott starts by noting the two basically different 
philosophical positions, that he calls Platonist, and non-Platonist.

Plato discussed  the imperfectness of our sensory perceptions, and compared them to the world seen by some hypothetical cavemen, who  have never been outside the cave, and can only perceive the world outside from the shadows they cast on the walls of the cave. From this example, he argued that 
there is a world outside, independent of our sensory perceptions, and this is the actual world, and 
what we perceive by our senses are only shadows.  Following this line of thought,   Mathematical forms, like natural numbers are a part of this  ideal world outside, and they have their own existence, independent of our perception. 

A natural extension of this viewpoint is the idea that numbers like  $7$, and $\pi$ were there even when mankind was not there.    Note that I chose the time   in the question  to be 100,000 years. It is much smaller than the age of the big bang, or the age of earth (about 4 billion years). By that time, most of the dinosaurs were long extinct, but humans were not clearly distinguishable from other apes.

The opposing position is that our  Mathematics is very much a result of human cultural evolution. And, further, mathematical forms are {\it made by people as we go along}, tailoring them to describe reality.  For a non-Platonist, there is no perfect circle, anywhere in the universe, and $\pi$ is merely a useful mental construct.

Abbott says that in his experience, about $80\%$ of mathematicians are Platonist, while engineers typically tend to be non-Platonists. Physicist, he says, are often `closet-non-Platonists':  in public, they side with the Platonist position, but are unsure of it, in their hearts. 

Regarding the unreasonable effectiveness, Abbott's  view is that mathematics is not very successful in most real-world problems, and the apparent effectiveness is a result of focussing only on the cases where it works. Mathematics is much less successful in describing biological systems, and even less in describing economic and social systems. We have cherry-picked a few successful cases, out of a large number of much less fortunate ones. Mathematics can appear to have an illusion of success, if we are preselecting the subset of problems for which we have found a way to apply mathematics.

About the Kepler problem, discussed by Wigner, he says that it is a self-selected example, and  relies on our fascination with squared numbers. Actual orbits are elliptical only to finite accuracy, and, in any case, the Newtonian theory is only an approximation to general relativity. While the elliptical orbit is a very good approximate description, {\it this is not an  example of Mathematics   miraculously  letting us arrive at the true nature of things}.  

The non-Platonist position is that mathematics is a product of human imagination.
All our physical laws are based on some simplifications/ idealizations/ approximations, and hence are  imperfect. Mathematics is a human invention for describing patterns and regularities.  It follows that mathematics is a useful tool for describing the regularities we see in the universe.

\section{My own position}

If Mathematics is independent of human existence, we can try to see what role Mathematics played in the world, when humans were not there,  as seen by other animals, say worms, frogs, birds, or even dogs, and monkeys. This is the reason for the choice of 100,000 years as the time in the question. In the case of bacteria or worms, it is hard to see what role Mathematics has in their world.  Birds have bigger brains, and  form bonds with mates.  Some bird-species are known to show disturbed behavior if some eggs are removed from their nest, which is evidence that they can distinguish between two or three eggs  in the nest, and thus can count up to three or four.  It is hard to imagine  Mathematics  playing any more significant  role than this  in the mental or worldly life of birds. 

In discussing the role of Mathematics in the world of other animals, it seems useful to  distinguish between different levels of mathematics. 

The zeroth level, that I  will call pre-Mathematics, is  the innate sense of numbers and shapes, that we share with other animals. This is the result of  evolution.  It helps animals move about in their environment, catch prey, or evade predators.  It is clearly very effective in this.  I think it is fair to say that Mathematics used by other mammals like dogs, and horses, even monkeys does not go substantially beyond this level. 

The first level will be the math that is expected to be known to students passing out of primary school. It consists of some familiarity with simple operations with integers, or fractional numbers, how to add them, or multiply, etc.. Not much more.

The second level, which I will call high-school level mathematics, involves the use of symbols, the notion of proof, and  abstractions like $\sqrt{2}$. This is the kind of math we teach children in high school, and this is all the math that even humans knew, even as late as  a few thousand years ago.  This is necessary for commerce and engineering.  In buying and selling, we (mostly) need to know only the addition and  multiplication of numbers. To make buildings that do not fall down, we need to use concepts from geometry, and strengths of materials, used in setting up buildings,  can be expressed in simple power-law relations between load and size etc. This uses level-2 math. This is also very effective, and perhaps it is not  surprizing, or `unreasonable'  that this is so. 

The third level, which I will call higher mathematics, is what is not covered in the first two levels.
Clearly, there are no sharp boundaries between these levels. For example, I chose not to include calculus in level-2, but could have. 

A biologist will note that there is no credible evidence of behaviour in non-human animals  showing that they have the  mental capability of dealing with any level-1 mathematics. So, I am not sure of what Mathematics we could be talking about in a world without humans \cite{klingons}.  If that sounds  too anthropo-centric, let me add that this is not  a question of humans versus non-humans.  Even amongst humans, mathematical ability is not very uniformly distributed. At a conservative estimate, about half of the population of high-school students, in all countries,  have serious difficulty with level-2 math.  I should add that this does not seem to seriously affect their  ability to enjoy life, or contribute to society.

\begin{figure}
\includegraphics[width=2.5cm,angle=0] {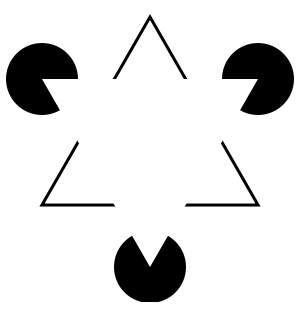}
\caption{The Kanisza triangle. We tend to see a well-formed inverted triangle, which is actually not there. Taken from  https://explorepsychology.wordpress.com/2011/11/25/illusions/ }
\end{figure}

Being able to detect patterns, approximate and idealize is a capability that we, as a species, have learnt through evolution.  This capability is rather basic, and an   example is  the  well-known visual illusion known as the Kanisza triangle ( see Figure 2).  It is clearly useful to be able to make sense out of the noisy and incomplete sensory data that we get. This ability  presumably also exists in other animals. Clearly, it helps in survival if you can detect a predator, partly hidden in grass. Seeing ellipses in planetary orbits is an example of the same idealization/ approximation/ filtering process.  

However, it would not be correct to equate this ability with Mathematics. It seems to me that `unreasonable effectiveness'  that Wigner is  talking about  is {\it mainly about mathematics of the third level.} 
In areas like  Biology, Psychology, even Geology, mathematics is not particularly useful.  So, let us agree that he is mainly thinking about Physics, even when he speaks of the 'Sciences'. 
Even in physics, in areas where one would expect  math to be effective, like predicting the motion of a cricket ball, you do not have to be a Tendulkar, or Kohli, to realize  that, in the real play, it is not. I find fully convincing  Abbott's  argument that  the "Effectiveness" of Mathematics is a result of the making the scope of regime of application very limited, only to questions where it is effective. Then Mathematics  {\it is} effective, but expectedly!

To come back to equation $s = \frac{1}{2} g t^2$ discussed earlier, one important {\it implicit assumption} in this equation is that distance moved may be represented as a real number. One may ask  if  the $15^{th}$ place in the decimal representation of distance moved expressed in cms, {\it exists}, in any real sense, for a macroscopic  body falling in air. Note that we are not talking of Planck length scales ($10^{-35}$ meters, where quantum gravity effects studied in string theory come into play), but this is  still about $10^{-5}$ of the size of an atom. A bit of thought will show that actually, even the concept of  center-of-mass is not well-defined at this level of precision. All the atoms in the ball jiggle, due to thermal motion, with amplitues of order $10^{-8}$ cm.   Also, some molecules would be getting  rubbed off, as the ball moves through the air, and I am not sure if their position should be included in the calculation of the center-of-mass. This again underlines the point that we simplify, idealize the actual problem, and describe the height of the falling ball  by a single {\it real} number,   and only then the simplified problem becomes tractable using some mathematical tools.

 That Mathematics deals only with a small set of possible scientific questions, was well-appreciated  by Wigner. He wrote, in the same article:\\
~\\
{\it "All the laws of nature are conditional statements that permit prediction of some future events ...( like position of the planets) on the basis of some aspects of the present state... As regards the present state of the world, such as the existence of the planet earth on which we live, and on which Galileo's experiments were performed, the existence of the sun, and all our surroundings, the laws of nature are entirely silent."}

On the meaning of `existence' in the   Platonic  world:  our conviction that a given table `exists' arises out our experience of  seeing it, and  feeling it by touch etc.. These direct sensory experiences  may  nowadays be augmented by more sensitive instruments  like microscopes,  x-ray cameras, or chemical sensors, or particle-colliders, if needed. It is the world of real objects like tables and rivers, and Higgs bosons, that Science deals with. Plato has used a very misleading analogy, and assigns cavemen's guess of what the  shadows might  be a higher level of existence  than to  the shadows themselves. We may even be willing to assign a higher status to the idea of a table, than to the table itself. But   does that  make the former  more real?

\section{Concluding remarks}

A discussion of the relationship between Mathematics and Physics, is not complete without some mention of the interactions between mathematicians and physicists ( in their professional capacities). Here,  I will like to retell my  favorite story, that I read first in an article by C. N. Yang ( unfortunately I have been  unable to track the original article).  It concerns a physicist who is travelling across USA, attending conferences, and giving lectures at different places.  He arrives in a small university town, checks into a hotel, and is walking up and down unfamiliar streets, with a bag of laundry, looking for a place where he can wash them. After a long walk, finally he finds a shop with a sign board ``Laundry done here".  He is much relieved, enters the shop, and puts his big bag of laundry on the counter.  On the other side,  is an oriental-looking man.  The man  seems mildly annoyed, looks up at the physicist, and asks,``What you want?". The physicist is a bit angry. He says,``I want my laundry done.''
``No laundry here.'' The physicist objects, and points to the sign-board, ``But see, it says right here, that  laundry is done  here." The man behind the counter smiles,"Ah, that! We make signs." 

I like the story because it captures the   frustration of  physicists in trying to get a mathematician  to help them in their work.  Quite often, a mathematician would  find the problem  the physicist wants to address uninteresting.  Of course, the reverse is also true. For example, a mathematician would worry about the existence and uniqueness of a solution, which a physicist is quite happy to take for granted.  Certainly, in most cases, the driving force behind a mathematician's work is not its usefulness to science.

To return to the  question we posed in the beginning of the article: `` Did $\pi$ exist 100,000 years ago?"  I have tried to convince you that the answer has to be `No'. A concept may be said to exist, after the first time someone thought of it, but  {\it even before that?}  What is true about $\pi$,  holds also for other more sophisticated mathematical constructs. All the mathematics we know is made by humans, and the same holds for mathematical concepts. For a more detailed discussion of this broader thesis from different perspectives, I can recommend   the collection of essays \cite{trick-or-truth}, available on the internet.
I will end by  quoting two sentences from  the essay by S. Wenmackers in the collection    \cite{wenmackers}:

{\it ``The fact that our so-called laws can be expressed with the help of mathematics should be telling, since it is {\rm our} (emphasis in original) science of patterns. When we open Galileo's proverbial book of nature, we find it is full of our own handwriting."}


\begin{thebibliography}{999}
\bibitem{crest} ``Like the crest of a peacock, like the gem, on the head of a snake, so is mathematics at the head of all knowledge.'',  translated from Vedanga Jyotisa (c. 500 BC)
\bibitem{klingons} Discussion about   math that is possibly used by the   klingons from another planet will be left for science-fiction fans.
\end{thebibliography}
\end{document}